\documentclass[11pt]{article}
\usepackage{longtable,supertabular}
\usepackage{amsmath}
\usepackage{amssymb}
\usepackage{latexsym}

\newcommand{\tabincell}[2]{\begin{tabular}{@{}#1@{}}#2\end{tabular}}

\title{\bf New Constructions of Subspace Codes Using Subsets of MRD codes in Several Blocks}
\author{Hao Chen, Xianmang He, Jian Weng and Liqing Xu
  \thanks{Hao Chen, Jian Weng and Liqing Xu are with the College of Information Science and Technology/Cyber Security, Jinan University, Guangzhou, Guangdong Province, 510632, China, haochen@jnu.edu.cn, cryptjweng@gmail.com, lqxu@fudan.edu.cn. Xianmang He is with the School of Network Security, Dongguan University of Technology, Guangdong Province, Dongguan, China, hexm@dgut.edu.cn. The research of Hao Chen and Liqing Xu  was supported by NSFC Grant 11531002. The research of Jian Weng was supported by NSFC Distinguishing Young Scholar Grant 61825203. The research of Xianmang He was supported by NSFC Grant 61672303,U1509213.}}

\begin{document}

\maketitle
\begin{abstract}
A basic problem for the constant dimension subspace coding is to determine the maximal possible size ${\bf A}_q(n,d,k)$ of a set of $k$-dimensional subspaces in ${\bf F}_q^n$ such that the subspace distance satisfies $d(U,V)=2k-2\dim(U \cap V) \geq d$  for any two different subspaces $U$ and $V$ in this set. We present two new constructions of constant dimension subspace codes using subsets of maximal rank-distance (MRD) codes in several blocks. This method is firstly applied to the linkage construction and secondly to arbitrary number of blocks of lifting MRD codes. In these two constructions, subsets of MRD codes with bounded ranks play an essential role. The Delsarte theorem of the rank distribution of MRD codes is an important ingredient to count codewords in our constructed constant dimension subspace codes. We give many new lower bounds for ${\bf A}_q(n,d,k)$. More than $110$ new constant dimension  subspace codes better than previously best known codes are constructed.\\
\end{abstract}

\section{Introduction}

Subspace coding was proposed by R. Koetter and F. R. Kschischang in \cite{KK}  to correct errors and erasures in random network coding (see \cite{EtzionVardy,Silva,KSK}). A set ${\bf C}$ of $M$ subspaces of the dimension $k$ in ${\bf F}_q^n$ is called a $(n, M, d, k)_q$ constant dimension subspace code (CDC) if  $d(U,V)=\dim(U+V)-\dim(U \cap V)=2k-2\dim(U \cap V) \geq d$ is satisfied for any given two distinct subspaces $U,V$ in ${\bf C}$. A main problem for subspace coding is to determine the maximal possible size ${\bf A}_q(n, d, k)$ of such a code for given parameters $n,d,k,q$. \\

Maximum rank-distance (MRD) codes have been widely used in the constructions of large constant dimension subspace codes. The rank metric on the space ${\bf M}_{m \times n}({\bf F}_q)$ of size $m \times n$ matrices over ${\bf F}_q$ is defined by the rank of matrices. That is the distance $d_r(A,B)$ is the rank of the matrix $A-B$. The minimum rank-distance of a code ${\bf M} \subset {\bf M}_{m \times n}({\bf F}_q)$ is defined as $$d_r({\bf M})=\min_{A\neq B} \{d_r(A,B), A \in {\bf M}, B\in {\bf M} \}$$ For a code ${\bf M}$ in ${\bf M}_{m \times n}({\bf F}_q)$ with the minimum rank distance $d_r({\bf M}) \geq d$, it is well-known that the number of codewords in ${\bf M}$ is upper bounded by $q^{\max\{m,n\}(\min\{m,n\}-d+1)}$ (see \cite{Delsarte,Gabidulin,Cruz}). A code attaining this bound is called a maximum rank-distance (MRD) code. The MRD code ${\bf Q}_{q,n,t}$ consists of ${\bf F}_q$ linear mappings on ${\bf F}_q^n \cong {\bf F}_{q^n}$ defined by $q$-polynomials $a_0x+a_1x^q+\cdots+a_ix^{q^i}+\cdots+a_tx^{q^t}$, where $a_t,\ldots,a_0 \in {\bf F}_{q^n}$ are arbitrary elements in ${\bf F}_{q^n}$.  The rank-distance of ${\bf Q}_{q,n,t}$ is $n-t$ since there are at most $q^t$ roots in ${\bf F}_{q^n}$ for each such $q$-polynomial. There are  $q^{n(t+1)}$ such $q$-polynomials in ${\bf Q}_{q,n, t}$ (see \cite{Gabidulin,Delsarte}). This kind of MRD codes have been used widely in previous constructions of constant dimension subspace codes (see \cite{Silberstein1,Silberstein2,Honold,Honold1,Skachek}).\\

In this paper firstly we give a parallel linkage construction based on the linkage construction proposed by Gluesing-Luerssen and Troha in \cite{Gluesing}. The basic idea is to use parallel versions of linkage and to give a suitable sufficient condition such that the subspace distance can be preserved for picking up subsets in these parallel blocks. This lead to some new record lower bounds which are better than the best known lower bounds in \cite{table}. The important ingredient is the Delsarte Theorem on the rank distribution of a MRD code.  The idea using matrices having lower and upper bounded ranks first appeared in \cite{Heinlein}.\\

Secondly we also give a construction of  constant dimension  subspace codes from several parallel versions of lifted MRD codes. The basic idea is as follows. If we only use elements $A_1,A_2,\ldots,A_s$ in a subset of the MRD code ${\bf Q}_{q,n,t}$ to construct dimension $n$ subspaces in ${\bf F}_q^{(s+1)n}$ spanned by the rows of the $n \times (s+1)n$ matrix  $(A_1,\ldots,I_n\ldots,A_s)$, then  for two such $n$-dimensional subspaces in ${\bf F}_q^{(s+1)n}$ the dimension of their intersection is at most $t$ since $A_1,\ldots,A_s$ are in the MRD code ${\bf Q}_{q,n,t}$. On the other hand some other $n$-dimensional subspaces in ${\bf F}_q^{(s+1)n}$ spanned by the rows of $(B_1,\ldots,I_n,\ldots,B_s)$, where $B_1,B_2,\ldots,B_s$ are  in the MRD code ${\bf Q}_{q,n,t}$, can be used to increase the size of constructed constant dimension subspace codes. Here we require that $I_n$ appears at different block positions.  We actually take $s+1$ subsets of $s+1$ parallel versions of lifted MRD codes. The key point here is to keep the subspace distances larger than or equal to $2(n-t)$ by a suitable sufficient condition on these $s+1$ subsets. By using this method  many new lower bound for  ${\bf A}_q((s+1)n,2(n-t), n), 2t \geq n$ are given with the help from the Delsarte Theorem about the rank distributions of the MRD code ${\bf Q}_{q,n,t}$. \\

In some cases new constant dimension subspace codes in the second construction are better. More than $110$ new constant dimension subspace codes better than \cite{table} are constructed. We give some examples in this paper and refer to the tables of  new constant dimension subspace codes to the full version \cite{CHWX} of this paper.\\

\section{Known results}

\subsection{Lifted MRD code}

Let $\displaystyle{n \choose k}_q=\prod_{i=0}^{k-1} \frac{q^{n-i}-1}{q^{k-i}-1}$ be the $q$-ary Gauss coefficient, which is the number of $k$-dimensional subspaces in ${\bf F}_q^n$. For any given MRD code ${\bf M}$ in ${\bf M}_{n \times n}({\bf F}_q)$ with the rank distance $d$, we have an $(2n, q^{n(n-d+1)}, 2d,n)_q$ CDC consisting of $q^{n(n-d+1)}$ subspaces of dimension $n$  in ${\bf F}_q^{2n}$ spanned  by the rows of $(I_n, A)$, where $A$ is an element in ${\bf M}$. Here $I_n$ is the $n \times n$ identity matrix. It is clear that for  $A$ and $B$, the subspaces $U_A$ and $U_B$ spanned by rows of $(I_n,A)$ and $(I_n,B)$ are the same if and only if $A=B$. The intersection $U_A \cap U_B$ is the set $\{ (\alpha,\alpha A)=(\beta, \beta B): \alpha (A-B)=0, \alpha \in {\bf F}_q^n\}$. Thus $\dim(U_A \cap U_B)  \leq n-d$. The distance of this CDC is $2d$. An CDC  constructed as above is called a lifted  MRD  code $C^{MRD}$, we refer to Proposition 4 in \cite{Silva} for the general form.\\

\subsection{Delsarte Theorem}

The rank distribution of a code ${\bf M}$ in ${\bf M}_{m \times n}({\bf F}_q)$ is defined by $A_i({\bf M})=|\{M \in {\bf M}, \operatorname{rank}(M)=i\}|$ for $i \in {\bf Z}^{+}$ (see \cite{Delsarte,Cruz}). The rank distribution of a MRD code can be determined from its parameters. We refer the following result to Theorem 5.6 in \cite{Delsarte} or Corollary 26 in \cite{Cruz}. The Delsarte Theorem is essential in this paper.\\

{\bf Theorem 2.1 (Delsarte 1978)} {\em Assume that ${\bf M} \subset {\bf M}_{n \times n}({\bf F}_q)$ is an MRD code with rank distance $d$, then its rank distribution is given by $$A_r({\bf M})=\displaystyle{n \choose r}_q \Sigma_{i=0}^{r-d} (-1)^i q^{\displaystyle{i \choose 2}} \displaystyle{r \choose i}_q (\frac{q^{n(n-d+1)}}{q^{n(n+i-r)}}-1).$$}

{\bf Corollary 2.1.} {\em Assume that ${\bf M} \subset {\bf M}_{n \times n}({\bf F}_q)$ is an MRD code with rank distance $d$, then $$A_d({\bf M})=(q^n-1)\displaystyle{n \choose d}_q .$$\\ $$A_{d+1}({\bf M})=\displaystyle{n \choose d+1}_q(q^{2n}-1-\frac{q^{d+1}-1}{q-1} (q^n-1)).$$\\ $$A_{d+2}({\bf M})=\displaystyle{n \choose d+2}_q (q^{3n}-1-\frac{q^{d+2}-1}{q-1}(q^{2n}-1)+q(\frac{q^{d+2}-1}{q^2-1})(\frac{q^{d+1}-1}{q-1})(q^n-1)).$$}\\

Let ${\bf Q}_{q, n, t, k} \subset {\bf Q}_{q,n,t}$ be the set of all $q$-polynomials in ${\bf Q}_{q,n,t}$ satisfying that the dimensions of kernels of the corresponding ${\bf F}_q$-linear mappings on ${\bf F}_{q^n}$ are bigger than or equal to $k$. It is clear that there is a filtration on ${\bf Q}_{q,n,t}$: ${\bf Q}_{q,n,t,t} \subset {\bf Q}_{q,n,t,t-1} \subset \cdots \subset {\bf Q}_{q,n,t,1} \subset {\bf Q}_{q,n,t,0}={\bf Q}_{q,n,t}$. Actually the cardinalities of these subsets in ${\bf Q}_{q,n,t}$ can be given from the Delsarte Theorem 2.1. We have the explicit formula for the cardinality of the space ${\bf Q}_{q,n,t,j}$ when $j \leq t$.\\

$$|{\bf Q}_{q,n,t,j}|=\Sigma_{i=n-t}^{n-j}A_i({\bf Q}_{q,n,t}).$$\\

\subsection{Previous constructions}

One upper bound is the anticode bound (see Theorem 5.2 in \cite{WXS2003} or Theorem 1 in \cite{EtzionVardy}) of CDC as follow. $${\bf A}_q(n, 2\delta, k) \leq \frac{\displaystyle{n \choose k-\delta+1}_q}{\displaystyle{k \choose k-\delta+1}_q}.$$  This showed that the ratio of this upper bound to the cardinality $|{\bf C}^{MRD}|$ depends on $q$ (see Lemma 9 on page 1008 of \cite{Silberstein2}). Hence a lower bound of ${\bf A}_q(n,d,k)$ should be compared with the size $|{\bf C}^{MRD}|$ of ${\bf C}^{MRD}$. The Johnson type bound (see Theorem 4 in \cite{EtzionVardy}) is the lower bound $${\bf A}_q(2n-1,2(n-t),n-1) \geq \frac{1}{q^n+1}{\bf A}_q(2n, 2(n-t),n).$$ This lower bound can be used to get some better constant dimension subspace codes in our construction.\\

We refer to some known results about general lower bounds for ${\bf A}_q(n,d,k)$ to  \cite{Silberstein1,Silberstein2,Silberstein3,Skachek}. Many CDCs from the multilevel construction based on echelon-Ferrers diagrams have been given. For example it was proved in \cite{Silberstein3} that when $q^2+q+1 \geq n-\frac{k^2+k-6}{2}$ and in some other cases (see \cite{Silberstein3})
\begin{displaymath}
{\bf A}_q(n,2(k-1), k)\geq q^{2(n-k)}+\Sigma_{j=3}^{k-1} q^{2(n-\Sigma_{i=j}^k i)}+\displaystyle{n-\frac{k^2+k-6}{2} \choose 2}_q.
\end{displaymath}
It was also  proved in \cite{Silberstein3} that
\begin{displaymath}
{\bf A}_q(n,4,k) \geq \Sigma_{i=1}^{[\frac{n-2}{k}]-1}(q^{(k-1)(n-ik)}+\frac{(q^{2(k-2)}-1)(q^{2(n-ik-1)}-1)}{(q^4-1)^2} q^{(k-3)(n-ik-2)+4}).
\end{displaymath}

If $n \geq 2k+2$ then $${\bf A}_q(n, 2, k) \geq \Sigma_{i=1}^{\lfloor\frac{n-2}{k}\rfloor-1}(q^{(k-1)(n-ik)}+\frac{(q^{2(k-2)}-1)(q^{2(n-ik-1)}-1)}{(q^4-1)^2}q^{(k-3)(n-ik-2)+4}).$$ This was proved in \cite{Silberstein3} Corollary 27.\\

The linkage construction in \cite{Gluesing} and the generalization in \cite{Heinlein} were used to give many presently best known lower bounds for constant dimension subspace codes with small parameters, we refer to \cite{table}.\\

\section{Parallel Linkage construction}

\subsection{General construction}

We recall some basic notations of linkage in \cite{Gluesing}. A set ${\bf U} \subset {\bf M}_{k\times n}({\bf F}_q)$ of $k \times n$ matrices over ${\bf F}_q$ is called a SC-representation of a set of $k$ dimensional subspaces in ${\bf F}_q^n$ if the rank of the matrices $U$ is $k$ for all $U \in {\bf U}$ and $\Im(U_1)\neq \Im(U_2)$ for all $U_1 \neq U_2$ in ${\bf U}$. Here $\Im(U)$ is the $k$ dimensional subspace spanned by the $k$ rows of $U$. The following Proposition 3.1 is a weaker version of the linkage construction in \cite{Gluesing}.\\

{\bf Proposition 3.1.} {\em  Let ${\bf U}$ be a SC-representation of a $(n_1, N_1, d_1,k)_q$ constant dimension subspace code and ${\bf Q} \subset {\bf M}_{k \times n_2}({\bf F}_q)$  be a code with rank distance $d_2$ and $N_2$ elements. Consider the set of $k$-dimensional subspaces in ${\bf F}_q^{n_1+n_2}$ defined by ${\bf C}=\{\Im(U, Q): U \in {\bf U}, Q \in {\bf Q}\}$. This is a $(n_1+n_2, N_1N_2, \min\{d_1,2d_2\}, k)_q$ constant dimension subspace code. Here $(U|Q)$ is a $k \times (n_1+n_2)$ matrix concatenated from $U$ and $Q$.}\\

{\bf Proof.} The intersection of two such $k$-dimensional subspaces  $W_1=\{x(U_1, Q_1): x \in {\bf F}_q^k\}$ and $W_2=\{y(U_2, Q_2): y \in {\bf F}_q^k\}$ in ${\bf F}_q^{n_1+n_2}$ is $$W_1 \cap W_2 =\{x(U_1, Q_1)=y(U_2, Q_2): x \in {\bf F}_q^k, y \in {\bf F}_q^k\}.$$ If $U_1 \neq U_2$, it is clear that the dimension of this subspace is smaller than or equal to the dimension of $\{xU_1=yU_2: x \in {\bf F}_q^k, y \in {\bf F}_q^k\}$. Then the subspace distance fulfills $d(W_1,W_2) \geq d_1$. If $U_1=U_2$, then for such $x$ and $y$ we have $x=y$ since $U_1=U_2$ is full rank. Then $\dim (W_1 \cap W_2) \leq \dim(\{x: x(Q_1-Q_2)=0\})=\dim(\ker(Q_1-Q_2))$. Then we have $d(W_1,W_2) \geq 2d_2$. The conclusion follows directly.\\

The following result is our parallel construction applied to the linkage construction.\\

{\bf Theorem 3.1 (Parallel linkage construction).} {\em Let ${\bf U}$ and ${\bf V}$ be  SC-representations of two  $(k+n, N_1, d, k)_q$ and $(n+k, N_2, d, k)_q$ constant dimension subspace codes satisfying that each $k \times (k+n)$ matrix in ${\bf U}$ is of the form $(U_1, U_2)$, where $U_1$ is a non-singular $k \times k$ matrix. We assume that $d \leq k$. Let ${\bf Q}_1 \subset {\bf M}_{k \times k}({\bf F}_q)$ be a code with rank distance $\frac{d}{2}$ and $N_3$ elements. Let ${\bf Q}_2 \subset {\bf M}_{k \times k}({\bf F}_q)$ be a code with rank distance $\frac{d}{2}$ and $N_4$ elements such that the rank of each element in ${\bf Q}_2$ is at most $k-\frac{d}{2}$. Then we have a $(k+n+k, N_1N_3+N_2N_4, d, k)_q$ constant dimension subspace code.}\\

{\bf Proof.} The code is defined by $${\bf C}=\{\Im(U_1, U_2, Q): (U_1, U_2) \in {\bf U}, {\bf Q} \in {\bf Q}_1\} \cup \{\Im(Q', V_1, V_2): Q' \in {\bf Q}_2, (V_1, V_2) \in {\bf V}\}.$$ From the proof of Proposition 3.1, the subspace distances of the two codes $${\bf W}_1=\{\Im(U_1, U_2, Q): (U_1, U_2) \in {\bf U}, {\bf Q} \in {\bf Q}_1\}$$ and $${\bf W}_2=\{\Im(Q', V_1, V_2): Q' \in {\bf Q}_2, (V_1, V_2) \in {\bf V}\}$$ are at least $d$. We only need to prove that the subspace distance of $W_1 \in {\bf W}_1$ and $W_2 \in {\bf W}_2$ is at least $d$. Thus these two codes are disjoint.\\

Consider $W_1 \cap W_2 =\{x(U_1, U_2, Q)=y(Q', V_1, V_2): (U_1, U_2) \in {\bf U}, (V_1, V_2) \\\in {\bf V}, Q\in {\bf Q}_1, Q'\in {\bf Q}_2, x, y \in {\bf F}_q^k\}$, then $xU_1=yQ'$. Since $rank(Q') \leq k-\frac{d}{2}$, $\dim(W_1 \cap W_2) \leq k-\frac{d}{2}$ because the matrix $U_1$ is a non-singular matrix and the dimension of the subspace $\{x: \exists y, xU_1=yQ'\}$ is at most the rank of the matrix $Q'$, that is $k-\frac{d}{2}$. Then $$d(W_1,W_2) \geq 2k -2(k-\frac{d}{2})=d.$$ The conclusion is proved.\\

\subsection{A new lower bound from parallel linkage}

Let $d$ and $k$ be two positive integers satisfying $d \leq k$ and $d$ be an even number. Set ${\bf U}=(I_k|Q)$ where $I_k$ is an identity matrix of size $k \times k$, where $Q$ is an arbitrary $q$-polynomial $a_{k-\frac{d}{2}} x^{q^{k-\frac{d}{2}}}+\cdots+a_1x^q+a_0x$, $a_i \in {\bf F}_{q^k}$. Let ${\bf V}$ be any $(2k, N, d, k)_q$ constant dimension subspace code. We set ${\bf Q}_1$ to be the MRD code ${\bf Q}_{q, k, k-\frac{d}{2}}$.  Let ${\bf Q}_2 \subset {\bf Q}_{q,k, k-\frac{d}{2}}$ be the set consisting of matrices of ranks $\frac{d}{2}, \frac{d}{2}+1, \ldots, k-\frac{d}{2}$, then $$|{\bf Q}_2|=\Sigma_{i=\frac{d}{2}}^{k-\frac{d}{2}} A_i ({\bf Q}_{q,k,k-\frac{d}{2}}).$$ ${\bf V}$ is an arbitrary $(2k, d, k)_q$ code. From Theorem 3.1 $${\bf A}_q(3k, d, k)  \geq q^{k(2k-d+2)}+(\Sigma_{i=\frac{d}{2}}^{k-\frac{d}{2}} A_i ({\bf Q}_{q,k,k-\frac{d}{2}})){\bf A}_q(2k,d, k).$$\\

When $k=6, d=6$, then $${\bf A}_q(18,6,6) \geq q^{48}+A_3({\bf Q}_{q,6,3}){\bf A}_q(12,6,6).$$ For example by using the lower bound ${\bf A}_2(12,6,6) \geq 16813481$ in \cite{table}, then ${\bf A}_2(18, 6, 6) \geq 282 952629488341$. The previously best known lower bound in \cite{table} is ${\bf A}_2(18,6,6) \geq 282 206 169223861$. The new lower bound ${\bf A}_2(18,6,6) \geq 282 957166112041$ by the parallel construction from MRD codes in Section 4 is better. If we use the improved lower bound ${\bf A}_2(12,6,6) \geq 16865101$ in Section 4 (or see \cite{XuChen}) we get the same lower bound ${\bf A}_2(18,6,6) \geq 282 957 166112041$ from our parallel linkage construction. \\

Let $h$ be a non-negative integer and $\phi: {\bf F}_{q^k} \longrightarrow {\bf F}_{q^{k+h}}$ be a $q$-linear embedding. Then $a_t \phi(x^{q^t})+a_{t-1}\phi(x^{q^{t-1}})+\cdots+a_1\phi(x^q)+a_0\phi(x)$ is a $q$-linear mapping from ${\bf F}_{q^k}$ to ${\bf F}_{q^{k+h}}$, where $a_i \in {\bf F}_{q^{k+h}}$ for $i=0,1,\ldots,t$. We denote the set of all such mappings as ${\bf Q}_{q, k\times (k+h), t}$. It is clear that the dimension of the kernel of any such mapping is at most $t$. Then ${\bf Q}_{q, k\times (k+h), t} \subset {\bf M}_{k \times (k+h)}({\bf F}_q)$ is a MRD code with rank distance $k-t$ and $q^{(k+h)(t+1)}$ elements. When $h=0$ we have the MRD code ${\bf Q}_{q,k,t}$.\\

Let $d$ and $k$ be two positive integers satisfying $d \leq k$ and $d$ be an even number. In Theorem 3.1 we set ${\bf U}=(I_k|Q)$ where $Q$ is an arbitrary element  in ${\bf Q}_{q, k\times(k+h),k-\frac{d}{2}}$. This is a $(2k+h, q^{(k+h)(k-\frac{d}{2}+1)}, d, k)_q$ code. Let ${\bf Q}_1$  be the MRD code ${\bf Q}_{q, k, k-\frac{d}{2}}$.  Let ${\bf Q}_2 \subset {\bf Q}_{q,k, k-\frac{d}{2}}$ be the set consisting of matrices of ranks $\frac{d}{2}, \frac{d}{2}+1, \ldots, k-\frac{d}{2}$. Thus $$|{\bf Q}_2|=\Sigma_{i=\frac{d}{2}}^{k-\frac{d}{2}} A_i ({\bf Q}_{q,k,k-\frac{d}{2}}).$$ ${\bf V}$ is an arbitrary $(2k+h,d, k)_q$ code. From Theorem 3.1 we have the following result.\\

{\bf Corollary 3.1.} {\em Let $h$ be a non-negative integer, $d$ and $k$ be positive integer. We assume that $d \leq k$ and $d$ is even. Then $${\bf A}_q(3k+h, d, k) \geq q^{(2k+h)(k-\frac{d}{2}+1)}+(\Sigma_{i=\frac{d}{2}}^{k-\frac{d}{2}} A_i({\bf Q}_{q,k,k-\frac{d}{2}})) {\bf A}_q(2k+h,d,k).$$}\\

When $k=6, d=6, h=1$ we get $${\bf A}_2(19,6,6) \geq 2^{52}+A_3({\bf Q}_{2,6,3}){\bf A}_2(13,6,6),$$ we have $A_3({\bf Q}_{2,6,3})=87885$ from the Delsarte Theorem. Then a new lower bound ${\bf A}_2(19,6,6) \geq 4527245732135821$ is proved, where the lower bound ${\bf A}_2(13,6,6) \geq 269057345$ in \cite{table} is used. The previously known best lower bound in \cite{table} is ${\bf A}_2(19,6,6) \geq 4515298730748862$.\\

The $63$ new constant dimension subspace codes better than \cite{table} by our parallel linkage construction Theorem 3.1 and Corollary 3.1 are listed in Table 1.\\

\section{Parallel construction using subsets of MRD codes in arbitrary number of blocks}

\subsection{General construction}

Similar to lifting the MRD code ${\bf Q}_{q,n,t}$  we have an  $((s+1)n, q^{sn(t+1)}, 2(n-t),n)_q$ CDC consisting $n$ dimensional subspace $U_{A_1,\ldots,A_s}^i$  in ${\bf F}_q^{(s+1)n}$ spanned by the rows of $n \times (s+1)n$ matrices $(A_1,\ldots,I_n,\ldots,A_s)$ with $q^{sn(t+1)}$ elements, where $A_1,A_2,\ldots,A_s$ takes all matrices in the MRD code ${\bf Q}_{q,n,t}$ and $I_n$ is at any position $i \in \{1,\ldots,s+1\}$. Then we consider other CDCs consisting of the  subspace $U_{B_1,\ldots,B_s}^j$ in ${\bf F}_q^{(s+1)n}$ spanned by the rows of the $ n \times(s+1)n$ matrices $(B_1,\ldots,I_n,,\ldots,B_s)$  where $B_1,\ldots,B_n$ are matrices from the MRD code ${\bf Q}_{q,n,t}$, $I_n$ is at a position $j \neq i$ in the set $\{1,\ldots,s+1\}$. \\

{\bf Proposition 4.1.} {\em The intersection $U_{A_1,\ldots,A_s}^i \cap U_{B_1,\ldots,B_s}^j$ is $$\{(\alpha A_1,\ldots, \alpha, \ldots, \alpha A_s)=(\beta B_1, \ldots,\beta, \ldots,\beta B_s): \exists \alpha \in {\bf F}_q^n, \exists \beta \in {\bf F}_q^n \}.$$ Then $\dim (U_{A_1,\ldots,A_s}^i  \cap U_{B_1,\ldots,B_s}^j) \leq n-rank(I_n-A_jB_i)=n-rank(I_n-B_iA_j).$ In particular  two subspaces in ${\bf F}_q^{(s+1)n}$ satisfy $U_{A_1,\ldots,A_s}^i=U_{B_1,\ldots,B_s}^j$ only if $A_jB_i=B_iA_j=I_n$.}\\

{\bf Theorem 4.1 (Parallel MRD construction with arbitrary number of blocks).} {\em If $2t \geq n$, then $${\bf A}_q((s+1)n, 2(n-t), n) \geq \Sigma_{j=0}^s q^{(s-j)n(t+1)}(\Sigma_{i=n-t}^t A_i({\bf Q}_{q,n,t}))^j.$$}\\

{\bf Proof 1.}  For the first block position of $I_n$ we take $n$-dimensional subspaces in ${\bf F}_q^{(s+1)n}$ spanned by rows of $(I_n, A_1^1,\ldots, A_s^1)$ where $A_1^1,\ldots,A_s^1$ are from the MRD code ${\bf Q}_{q,n,t}$. There are $q^{sn(t+1)}$ such subspaces. For the second block position of $I_n$ we take $n$-dimensional subspaces in ${\bf F}_q^{(s+1)n}$ spanned by rows of $( A_1^2,I_n,\ldots, A_s^2)$ where $A_1^2,\ldots,A_s^2$ are from the MRD code ${\bf Q}_{q,n,t}$ and $A_1^2 \in {\bf Q}_{q,n,t,n-t}$. There are $q^{(s-1)n(t+1)}|{\bf Q}_{q,n,t,n-t}|=q^{(s-1)n(t+1)}(\Sigma_{i=n-t}^t A_i({\bf Q}_{q,n,t})) $ such subspaces. For the third block position of $I_n$ we take $n$-dimensional subspaces in ${\bf F}_q^{(s+1)n}$ spanned by rows of $( A_1^3,A_2^3, I_n,\ldots, A_s^3)$ where $A_1^3,\ldots,A_s^3$ are from the MRD code ${\bf Q}_{q,n,t}$ and $A_1^3 \in {\bf Q}_{q,n,t,n-t}, A_2^3 \in {\bf Q}_{q,n,t,n-t}$. There are $q^{(s-2)n(t+1)}|{\bf Q}_{q,n,t,n-t}|^2=q^{(s-2)n(t+1)}(\Sigma_{i=n-t}^t  A_i({\bf Q}_{q,n,t}))^2$ such subspaces. We continue this process. All these subspaces in ${\bf F}_q^{n(s+1)}$ are different from Proposition 4.1.\\

For any fixed block position $j$ of $I_n$, the dimension of the intersection of two different subspaces is at most $t$ since $A_1^j,\ldots,A_s^j$ are in the MRD code ${\bf Q}_{q,n,t}$. For different block positions $j>i$ in the set $\{1,\ldots,s\}$, the dimension of the intersection of two different subspaces is at most $\dim(\ker(I_n-A_j^iA_i^j))$ from Proposition 4.1. Since the dimension of the space $$\ker(I_n-A_j^iA_i^j)=\{x:x=xA_j^iA_i^j\}$$ is at most $t$ from the fact $\dim(\ker(A_i^j)) \geq n-t$, we get the conclusion.\\

{\bf Proof 2.}  We use the same matrices $A_i^j$ as in the Proof 1. First of all these subspaces are different since $A_i^j$'s  with $j>i$ are singular matrices. For two block position indices $i<j$, the intersection of two $n$ dimensional subspaces in ${\bf F}_q^{(s+1)n}$ spanned by rows is of the following form $$\{\beta A_i^j: \beta \in {\bf F}_q^n \}.$$ Since the dimension of the kernel of $A_i^j$ is larger than or equal to $n-t$, the dimension of the space of all possible $\alpha$'s is at most $t$. The conclusion is proved.\\

\subsection{Some new lower bounds}

In the case $s=1$ we get the following result.\\

{\bf Corollary 4.1.} {\em If $2t \geq n$, we have ${\bf A}_q(2n, 2(n-t),n) \geq q^{n(t+1)}+\Sigma_{i=n-t}^t A_i({\bf Q}_{q,n,t})$.}\\

We list all $42$ improvements on \cite{table} in Table 2 of the full version \cite{CHWX} of this paper.\\

{\bf Corollary 4.2 (combining with the Johnson type bound).} {\em  If $2t \geq n$, we have ${\bf A}_q(2n-1, 2(n-t),n-1) \geq \frac{1}{q^n+1}(q^{n(t+1)}+\Sigma_{i=n-t}^t A_i({\bf Q}_{q,n,t}))$.}\\

We refer to Table 3 for $7$ new constant dimension subspace codes from Corollary 4.2 applied to parameters $n=9, t=6$.\\

From Theorem 4.1 we have the following Corollary 3.2 immediately.\\

{\bf Corollary 4.3.} {\em If $2t \geq n$ then  ${\bf A}_q(3n, 2(n-t),n) \geq q^{2n(t+1)}+\\q^{n(t+1)}(\Sigma_{i=n-t}^t A_i({\bf Q}_{q,n,t}))+(\Sigma_{i=n-t}^t A_i({\bf Q}_{q,n,t}))^2$.}\\

For example in the case $n=2k,t=k$ we have the following lower bound for ${\bf A}_q(6k,2k,2k)$.\\

{\bf Corollary 4.4.} {\em ${\bf A}_q(6k, 2k,2k) \geq q^{4k(k+1)}+q^{2k(k+1)} (q^{2k}-1)\prod_{i=0}^{k-1}\frac{q^{2k-i}-1}{q^{k-i}-1}+ ((q^{2k}-1)\prod_{i=0}^{k-1}\frac{q^{2k-i}-1}{q^{k-i}-1})^2$.}\\

We refer to Table 4 for $21$ new better constant dimension subspace codes in the case $s=2$.\\

From Theorem 4.1 the following lower bound can be proved for the case $s=3$.\\

{\bf Corollary 4.5.} {\em If $2t \geq n$ then  ${\bf A}_q(4n, 2(n-t),n) \geq q^{3n(t+1)}+\\q^{2n(t+1)}(\Sigma_{i=n-t}^t A_i({\bf Q}_{q,n,t}))+q^{n(t+1)}(\Sigma_{i=n-t}^t A_i({\bf Q}_{q,n,t}))^2+(\Sigma_{i=n-t}^t A_i({\bf Q}_{q,n,t}))^3$.}\\

In the case $n=2k,t=k$ we have the following lower bound from the Delsarte Theorem.\\

{\bf Corollary 4.6.} {\em ${\bf A}_q(8k, 2k,2k) \geq q^{6k(k+1)}+q^{4k(k+1)} (q^{2k}-1)\prod_{i=0}^{k-1}\frac{q^{2k-i}-1}{q^{k-i}-1}+ q^{2k(k+1)}((q^{2k}-1)\prod_{i=0}^{k-1}\frac{q^{2k-i}-1}{q^{k-i}-1})^2+((q^{2k}-1)\prod_{i=0}^{k-1}\frac{q^{2k-i}-1}{q^{k-i}-1})^3$.}\\

We list some lower bound for ${\bf A}_q(20,4,5)$ and ${\bf A}_q(24,6,6)$ in Table 5. No entries in \cite{table} can be compared with these lower bounds.\\

\section{Conclusion}

In this paper two parallel constructions of constant dimension subspace codes based on the linkage construction and arbitrary number of lifted MRD codes are given. Many new lower bounds on ${\bf A}_q(n,d,k)$ were proved from these parallel constructions using subsets of MRD codes in several blocks. The novelty of this paper is the using subsets counted by the Delsarte Theorem in several parallel blocks of lifted MRD codes. From Tables 1-5 in the full version \cite{CHWX} more than $110$ new constant dimension subspace codes better than \cite{table} have been constructed from our new parallel constructions. \\

{\bf Added Notes.} Some results in this paper have been extended in \cite{Cossidente,Heinlein1,LCF}.

{\bf Acknowledgement.} We are grateful to the first referee for his/her careful reading of the paper and very helpful comments. We thank the Associate Editor and the second referee for their suggestions.\\

\appendix
\section{Appendix}
\begin{longtable}{|l@{\extracolsep{\fill}}|l|l|l||}
 \caption{\label{tab:A-q-5-4}New subspace codes from parallel linkage}\\
\hline
${\bf A}_q(n,d,k)$　&　New 　&　Old　 \\ \hline  \hline \endfirsthead
\multicolumn{4}{r}{continued table} \\ \hline
${\bf A}_q(n,d,k)$ & New & Old \\ \hline \endhead
${\bf A}_2(15,4,5)$ & 1252409384941   &1235787711790     \\ \hline
${\bf A}_3(15,4,5)$ & 12399152701973746721   &12394544365887696067   \\ \hline
${\bf A}_4(15,4,5)$ &\tabincell{l}{1215514411297742359058971}   &     \tabincell{l}{1215478900794081741379237}  \\ \hline
${\bf A}_5(15,4,5)$ &\tabincell{l}{9113715532358125452199289569}   &     \tabincell{l}{9113676963739967346201192181}   \\ \hline
${\bf A}_7(15,4,5)$ &\tabincell{l}{6369953433032799634940814403\\550401}   &        \tabincell{l}{6369951878418978850938882154\\998943}  \\ \hline
${\bf A}_8(15,4,5)$ &\tabincell{l}{1329603936275508854413747923\\192211831}   &     \tabincell{l}{1329603830010446369320349184\\800629897}  \\ \hline
${\bf A}_9(15,4,5)$ &\tabincell{l}{1478344516592412811422378899\\08886770241}   &   \tabincell{l}{1478344472192502033634129606\\95716746417}   \\ \hline
${\bf A}_2(16,4,5)$ & 20021868703021   &19772603404689     \\ \hline
${\bf A}_3(16,4,5)$ & 1004246333824396831601   &1003958093636913086356   \\ \hline
${\bf A}_4(16,4,5)$ &\tabincell{l}{311169775104436392108967291}   &     \tabincell{l}{311162598603284926601722789}  \\ \hline
${\bf A}_5(16,4,5)$ &\tabincell{l}{569606782898065136733586366\\0369}   &             \tabincell{l}{569604810233747959140019927\\4056}   \\ \hline
${\bf A}_7(16,4,5)$ &\tabincell{l}{152942576811121287704054973\\91944583201}   &     \tabincell{l}{152942544600839682211042601\\25199891012}  \\ \hline
${\bf A}_8(16,4,5)$ &\tabincell{l}{544605767017671083057925917\\9780851932151}   &    \tabincell{l}{544605728772278832873615029\\1737261978761}  \\ \hline
${\bf A}_9(16,4,5)$ &\tabincell{l}{969941834171302637911606992\\269353378624481}   &  \tabincell{l}{969941808205500584267352435\\307639908204424}   \\ \hline

${\bf A}_2(17,4,5)$ & 320365633119931   &316361655057323     \\ \hline
${\bf A}_3(17,4,5)$ & 81343951054914823057601   &81320605584592333256896   \\ \hline
${\bf A}_4(17,4,5)$ &\tabincell{l}{796594623030380982288285895\\51}   &                    \tabincell{l}{796576252424409420394019074\\93}  \\ \hline
${\bf A}_5(17,4,5)$ &\tabincell{l}{356004239278632417623139424\\6966849}   &               \tabincell{l}{356003006396092474470158189\\5765556}   \\ \hline
${\bf A}_7(17,4,5)$ &\tabincell{l}{367215126923122857711324989\\25095511388801}   &       \tabincell{l}{367215049586616076988713969\\88494253570488}  \\ \hline
${\bf A}_8(17,4,5)$ &\tabincell{l}{223070522170400489301792865\\41042123379285951}   &     \tabincell{l}{223070506505125409945032726\\04052070322817161}  \\ \hline
${\bf A}_9(17,4,5)$ &\tabincell{l}{636378837399770196848173247\\3171613640616521601}   &   \tabincell{l}{636378820363628933337809933\\8862136580032063628}   \\ \hline

${\bf A}_2(18,4,5)$ & 5125557140935621   &5061786480788587     \\ \hline
${\bf A}_3(18,4,5)$ & 6586984882892620375466801   &6586969052351977742082856   \\ \hline
${\bf A}_4(18,4,5)$ &\tabincell{l}{203923535631374864844098226\\92731}   &                  \tabincell{l}{203923520620648811617657149\\36261}  \\ \hline
${\bf A}_5(18,4,5)$ &\tabincell{l}{222501880086200341340588393\\7426160369}   &             \tabincell{l}{222501878997557796543846638\\9564044756}   \\ \hline
${\bf A}_7(18,4,5)$ &\tabincell{l}{881683334130185794241976851\\18451711404973601}   &     \tabincell{l}{881683334057465200849902241\\56399442519707072}  \\ \hline
${\bf A}_8(18,4,5)$ &\tabincell{l}{913696794659942511668319032\\60963819527758453751}&    \tabincell{l}{913696794644993679134854045\\86035285169051407881}  \\ \hline
${\bf A}_9(18,4,5)$ &\tabincell{l}{417528144042218931834576543\\49776548099387974166561}   &\tabincell{l}{417528144040576943162937097\\62272976069664226251756}   \\ \hline

${\bf A}_2(19,4,5)$ &82000714657355896 & 80988583692738669   \\ \hline
${\bf A}_3(19,4,5)$ &533545775512317389092508801&    533544493240510197079493944   \\ \hline
${\bf A}_4(19,4,5)$ &\tabincell{l}{52204425121630728423907635\\42302191}   &            \tabincell{l}{52204421278886095774120022\\68988497}  \\ \hline
${\bf A}_5(19,4,5)$ &\tabincell{l}{13906367505387518067957491\\07370809466849}   &        \tabincell{l}{13906367437347362283990414\\91814662484526}   \\ \hline
${\bf A}_7(19,4,5)$ &\tabincell{l}{21169216852465760915956323\\5358302246119908739201}&   \tabincell{l}{21169216850719739472406152\\8199514099366794683044}  \\ \hline
${\bf A}_8(19,4,5)$ &\tabincell{l}{37425020709271245277558484\\3883548745445452491759551}&  \tabincell{l}{37425020708658941097363621\\7184400516522331058635329}  \\ \hline
${\bf A}_9(19,4,5)$&
\tabincell{l}{273940215306099841176451031\\332562928972470621968108481}&
\tabincell{l}{273940215305022532409203029\\750272995891203951349923082}\\
\hline

${\bf A}_2(18,4,6)$&1321065731337118327 & 1301902384896972957     \\ \hline
${\bf A}_3(18,4,6)$ &43241984500836016475467263377 &  43225562953761729683056546744     \\ \hline
${\bf A}_4(18,4,6)$ &\tabincell{l}{133649773466469722107967099790\\3343231} &    \tabincell{l}{13364584050324721907495003191\\15666769}    \\ \hline
${\bf A}_5(18,4,6)$ &\tabincell{l}{869154313455571766784982919397\\200744721649} & \tabincell{l}{86915062398553386450111346455\\8715816063570}  \\ \hline
${\bf A}_7(18,4,6)$ &\tabincell{l}{508273121397713162379788076978\\860454942323352347617}&\tabincell{l}{50827299725042503817812207998\\9337055565420133852250}\\ \hline
${\bf A}_8(18,4,6)$ &\tabincell{l}{153292907353372012238179216797\\2707130036770914661707007} & \tabincell{l}{15329289509595962385976540015\\68049806785911717931336256}    \\  \hline
${\bf A}_9(18,4,6)$ &\tabincell{l}{179732185298838953062296457191\\2268841144546741298887430561} & \tabincell{l}{17973217989922197607083648785\\65965820546280222286535998208}    \\ \hline

${\bf A}_2(19,4,6)$ & 42208289248791279191  & 41660876316712223851    \\ \hline
${\bf A}_3(19,4,6)$ &\tabincell{l}{10503812381857770555608261193\\203}  &\tabincell{l}{10503811797764100313173626438\\410}  \\ \hline
${\bf A}_4(19,4,6)$ &\tabincell{l}{13685334079363182818700343352459\\15778619}   &           \tabincell{l}{1368533406753251523327488538756\\265820613}  \\ \hline
${\bf A}_5(19,4,6)$ &\tabincell{l}{27160957000408320819947997516309\\16148882970869}   &   \tabincell{l}{2716095699954793272557435557305\\913288978107256}   \\ \hline
${\bf A}_7(19,4,6)$ &\tabincell{l}{85425442647896740463687280455997\\49570456742066054954023}   &  \tabincell{l}{8542544264787894328644239651424\\668612867543534298440406}  \\ \hline
${\bf A}_8(19,4,6)$ &\tabincell{l}{50231015865045467079871255733225\\534110744159933103898225143}  &   \tabincell{l}{5023101586504404954636792632338\\1856072893849422409772176905}  \\ \hline
${\bf A}_9(19,4,6)$ &\tabincell{l}{10613005490869209759549527518453\\4478863507421929255303118181289}&     \tabincell{l}{10613005490869158463473211884225\\8778340200008478852505231237828\\}   \\ \hline

${\bf A}_2(18,6,6)$  & 282952629488341   &   282206169223861  \\
\hline
${\bf A}_3(18,6,6)$  & 79773409456539341408321 &   7977052899429695519499   \\
\hline
${\bf A}_4(18,6,6)$  &  79228596836450068221001288411 & 79228465213535437618551984193    \\
\hline
${\bf A}_5(18,6,6)$ &\tabincell{l}{355271606149018080912048635023\\7569} &              \tabincell{l}{355271549860537797212597654883\\4375} \\
\hline
${\bf A}_7(18,6,6)$ & \tabincell{l}{367033693031672275575986481885\\27350390401} &       \tabincell{l}{367033691269048237620480816438\\30813838569}    \\
\hline
${\bf A}_8(18,6,6)$ & \tabincell{l}{223007453917574042420344146217\\65810585581431}&     \tabincell{l}{223007453646901902254328287720\\81255730905601} \\
\hline
${\bf A}_9(18,6,6)$ & \tabincell{l}{636268545986545103814893081316\\5508870317893121} &  \tabincell{l}{636268545755947049961803989258\\2186036406787787}    \\
\hline

${\bf A}_2(19,6,6)$ & 4527245732135821    & 4515298730748862   \\ \hline
${\bf A}_3(19,6,6)$ & 6461646166374500995275281    &  6461417369472937542117973  \\ \hline
${\bf A}_4(19,6,6)$ & \tabincell{l}{202825207901329766842749685\\10491}   &               \tabincell{l}{20282487415579548140041494\\597697}  \\ \hline
${\bf A}_5(19,6,6)$ & \tabincell{l}{222044753843136427265160012\\6040019569}   &          \tabincell{l}{22204471885174522176980972\\29003922001}   \\ \hline
${\bf A}_7(19,6,6)$ & \tabincell{l}{881247896969045133932229870\\51228915467935201}   &   \tabincell{l}{88124789274625593704300569\\118173789808207533}  \\ \hline
${\bf A}_8(19,6,6)$ & \tabincell{l}{913438531246383277768338625\\08434972183748561271}  & \tabincell{l}{91343853013939625003475366\\792854319199699599873}  \\ \hline
${\bf A}_9(19,6,6)$ & \tabincell{l}{417455793021772242613440430\\29615239593994446168481}&\tabincell{l}{41745579287064298367037383\\503578317354686374220297}   \\ \hline
\end{longtable}

\setlength{\LTleft}{0pt} \setlength{\LTright}{0pt} 

\begin{longtable}{|l@{\extracolsep{\fill}}|l|l|l||}

 \caption{\label{tab:test}New lower bounds on ${\bf A}_q(2n,2(n-t),n)$  in the case $s=1$}\\

\hline

${\bf A}_q(n,d,k)$　&　New 　&　Old　 \\ \hline  \hline \endfirsthead

\multicolumn{4}{r}{continued table} \\ \hline

${\bf A}_q(n,d,k)$&　New 　&　Old   \\ \hline  \hline \endhead

\hline
${\bf A}_2(12,6,6)$  & 16865101  &  16813481   \\
\hline
${\bf A}_3(12,6,6)$  & 282454201121 &   282444003514  \\
\hline
${\bf A}_4(12,6,6)$  & 281476519727131  &  281476052114497  \\
\hline
${\bf A}_5(12,6,6)$  & 59604684750269569  &  59604675306650126  \\
\hline
${\bf A}_7(12,6,6)$  & 191581237048517640001 & 191581236128477745586  \\
\hline
${\bf A}_8(12,6,6)$  & 4722366523787007379831  &  4722366518055302169089 \\
\hline
${\bf A}_9(12,6,6)$  & 79766443311676870053761  &  79766443282767710316742 \\
\hline
${\bf A}_2(14,6,7)$  & 34532238023  &  34432090228 \\
\hline
${\bf A}_3(14,6,7)$  & 50035894106387201 &  50031545103789355 \\
\hline
${\bf A}_4(14,6,7)$  & 1180598085852241507903  & 1180591620717679804753  \\
\hline
${\bf A}_5(14,6,7)$  & 2910384996920980634798249  & 2910383045673376465235151 \\
\hline
${\bf A}_7(14,6,7)$  & 378818703472375564718065717033 & 378818692265664782360946466387  \\
\hline
${\bf A}_8(14,6,7)$  & \tabincell{l}{40564819558769908757687294403071}  &    \tabincell{l}{40564819207303340852292565865025} \\
\hline
${\bf A}_9(14,6,7)$  & \tabincell{l}{250315551236152487860737633625\\4513}  & \tabincell{l}{250315550499324160133844882159\\4903}  \\
\hline
${\bf A}_2(16,8,8)$  & 1099562828461  & 1099528467457  \\
\hline
${\bf A}_3(16,8,8)$  & 12157665957047665121 &  12157665459056935444 \\
\hline
${\bf A}_4(16,8,8)$  & 1208925820022362618115611  &  1208925819614629174771969 \\
\hline
${\bf A}_5(16,8,8)$  & 9094947017807612368002449569  & 9094947017729282379150781876  \\
\hline
${\bf A}_8(16,8,8)$  & \tabincell{l}{13292279957849213674394203378\\73299831}  &  \tabincell{l}{1329227995784915872903807060\\297125889}  \\
\hline
${\bf A}_9(16,8,8)$  & \tabincell{l}{14780882941434601431198749296\\8729104321}  &\tabincell{l}{1478088294143459233160832102\\06426350884} \\
\hline
${\bf A}_2(16,6,8)$  & 282927683836351  &  282065502894292 \\
\hline
${\bf A}_3(16,6,8)$  & 79773403858211367304001 & 79766443077154959293127  \\
\hline
${\bf A}_4(16,6,8)$  & 79228596795209597286010744831  &  79228162514264619069883417872 \\
\hline
${\bf A}_5(16,6,8)$  &\tabincell{l}{355271606144635047856413687678\\1249}  &         \tabincell{l}{355271367880050098896030301250\\3775}  \\
\hline
${\bf A}_7(16,6,8)$  &\tabincell{l}{367033693031655064026816246271\\51289328001} &   \tabincell{l}{367033682172941254414217922688\\54792155015} \\
\hline
${\bf A}_8(16,6,8)$  & \tabincell{l}{22300745391757287672361562599\\998342819479551}  &\tabincell{l}{2230074519853062314154044063917\\0885812523072}  \\
\hline
${\bf A}_9(16,6,8)$  &
\tabincell{l}{63626854598654462048615260385\\54759414900421761}&
\tabincell{l}{63626854411359423584749085289\\81840165075519087}\\
\hline
${\bf A}_2(18,8,9)$  & 18015215398068295  &
                       18014674602898481 \\ \hline
${\bf A}_3(18,8,9)$  & 58149739380417667198523945
                     & 58149737003040060077869735  \\ \hline
${\bf A}_4(18,8,9)$ &
\tabincell{l}{32451855376784298642321288625\\1071}  &
\tabincell{l}{32451855365842672678322474110\\1633}  \\ \hline
${\bf A}_5(18,8,9)$  &
\tabincell{l}{555111512317358783579601168291\\64048249}           &\tabincell{l}{55111512312578270211815872192\\48047001}   \\ \hline
${\bf A}_7(18,8,9)$  &
\tabincell{l}{431811456739659181762301609528\\5299264536325745} &  \tabincell{l}{431811456739643656403529309770\\9356501432634891}  \\\hline
${\bf A}_8(18,8,9)$ &
\tabincell{l}{584600654932363583793403430292\\3933590182378512895}& \tabincell{l}{584600654932361167281473933086\\5150093023313396225} \\ \hline
${\bf A}_9(18,8,9)$ &
\tabincell{l}{3381391913522728424620280247018\\514713413256655866641}& \tabincell{l}{3381391913522726342930221472392\\241320293166235632813}\\ \hline
${\bf A}_2(18,6,9)$  & 9271545156551861247
                     & 9242714023345881465  \\ \hline
${\bf A}_3(18,6,9)$  &
\tabincell{l}{1144661280188113228748844786839} &
\tabincell{l}{1144561273430987589803690699062} \\ \hline
${\bf A}_4(18,6,9)$  &
\tabincell{l}{8507105814618280327650335111984\\8669183}  &                        \tabincell{l}{8507059173023462058821041620363\\9381057} \\ \hline
${\bf A}_5(18,6,9)$  &
\tabincell{l}{1084202899657109779066908450170\\97020371093749}  &                 \tabincell{l}{1084202172485504434152971953749\\65698255868876} \\ \hline
${\bf A}_7(18,6,9)$  &
\tabincell{l}{1742515033889755513188849225993\\69466772818993479502027} &         \tabincell{l}{1742514982336908143055132035255\\56311342652949519875530}  \\ \hline
${\bf A}_8(18,6,9)$  &
\tabincell{l}{7846377237219197911383816346352\\35733830468771226268467199}  &     \tabincell{l}{7846377169233350954794740024195\\11965161887731769327059457} \\\hline
${\bf A}_9(18,6,9)$  &
\tabincell{l}{1310020512493866339206870302329\\188713348371417431388541027913}  & \tabincell{l}{1310020508637620352391208118240\\901618983186587598009041655712}  \\
\hline
\end{longtable}

\begin{table}[hb]
\caption{New lower bounds from the Johnson type bound}\label{table-s-1-joh-type}
\begin{tabular}{|c|l|l|}\hline
${\bf A}_q(n,d,k)$　&　New 　&　Old   \\
\hline
${\bf A}_2(17,6,8)$ & 18073187439672244 &  18052309715589680\\
\hline
${\bf A}_3(17,6,8)$  &  58151863451946414791142287  &  58149737004893178906982592 \\
\hline
${\bf A}_4(17,6,8)$ & \tabincell{l}{32451909495196476483054550389\\9935}                 & \tabincell{l}{324518553658445173598894784069\\722}  \\
\hline
${\bf A}_5(17,6,8)$ &\tabincell{l}{555111600407300798344248374232\\36913732}&              \tabincell{l}{555111512312578503042579744633\\91093912} \\
\hline
${\bf A}_7(17,6,8)$ &\tabincell{l}{431811458814229328190145779776\\0474522447137650}  &    \tabincell{l}{431811456739643656513972079940\\3479106597531752} \\
\hline
${\bf A}_8(17,6,8)$ &\tabincell{l}{584600655642087187407545566975\\9065390165175356426}  & \tabincell{l}{584600654932361167289396749397\\0879175905108820562}\\
\hline
${\bf A}_9(17,6,8)$ &\tabincell{l}{338139191474840770349258063849\\2271571254198293516660}&\tabincell{l}{338139191352272634293365515622\\1409046767887551853552}\\ \hline
\end{tabular}
\end{table}

\begin{table}[hb]
\centering
\caption{New lower bounds on ${\bf A}_q(3n.2(n-t),n)$ in the case $s=2$} \label{table-bounds-s-2}
\begin{tabular}{|c|l|l|}\hline
${\bf A}_q(n,d,k)$ &  New  &  Code  \\
\hline
${\bf A}_2(18,6,6)$  & 282957166112041   &   282206169223861  \\
\hline
${\bf A}_3(18,6,6)$  & 79773409708059646924801 &   79770528994296955194991   \\
\hline
${\bf A}_4(18,6,6)$  &  79228596837171602219181433561 & 79228465213535437618551984193    \\
\hline
${\bf A}_5(18,6,6)$ & \tabincell{l}{355271606149055831666451347994\\5761} &              \tabincell{l}{355271549860537803173065185548\\4501} \\
\hline
${\bf A}_7(18,6,6)$ & \tabincell{l}{367033693031672327723398953389\\21195414401} &       \tabincell{l}{367033691269048247553967909699\\87924701979}    \\
\hline
${\bf A}_8(18,6,6)$ & \tabincell{l}{223007453917574044765606725592\\19358376203601}&     \tabincell{l}{223007453646901902254328287720\\81255730905601} \\
\hline
${\bf A}_9(18,6,6)$ & \tabincell{l}{636268545986545104493692756885\\8327487086310721} &  \tabincell{l}{636268545755947014524069928399\\2755392451222033}    \\

\hline

${\bf A}_2(18,4,6)$ &  1321055665352277121 & 1301902384896972957     \\
\hline
${\bf A}_3(18,4,6)$   &  43241984454039791949376848001 &  43225562953761729683056546744     \\
\hline
${\bf A}_4(18,4,6)$  & \tabincell{l}{13364977346615645679038498701\\19608321} &        \tabincell{l}{13364584050324721907495003191\\15666769}    \\
\hline
${\bf A}_5(18,4,6)$  & \tabincell{l}{86915431345555286301049529274\\6726010500001} &    \tabincell{l}{8691506239855338472183793783\\22608115640776 }    \\
\hline
${\bf A}_7(18,4,6)$  & \tabincell{l}{50827312139771315191417379894\\7508628845999547723521} &         \tabincell{l}{508272997250425080540503340954\\642021097480629123655}    \\
\hline
${\bf A}_8(18,4,6)$  & \tabincell{l}{15329290735337201203431548481\\57539946320174365857546241} &     \tabincell{l}{153292895095959623859765400156\\8049806785911717931405888}    \\
\hline
${\bf A}_9(18,4,6)$  & \tabincell{l}{17973218529883895304078740000\\31880315113074804045244546241} &  \tabincell{l}{179732179899221976044864635882\\5022918933920879979928528654}    \\
\hline
${\bf A}_2(15,4,5)$ &  1252379805361  & 1235787711790     \\
\hline
${\bf A}_3(15,4,5)$ &  12399152568347096641 &  12394544365887696067     \\
\hline
${\bf A}_4(15,4,5)$ &  1215514411238392851780481 &  1215478900794081741379237   \\
\hline
${\bf A}_5(15,4,5)$ &  9113715532351043940956916001 &  9113676963739967346201192181      \\
\hline
${\bf A}_7(15,4,5)$ &  \tabincell{l}{636995343303278946060145826616\\9601} &       \tabincell{l}{636995187841897885093888215499\\8943}     \\
\hline
${\bf A}_8(15,4,5)$ &  \tabincell{l}{132960393627550866960611827601\\3276161} &    \tabincell{l}{132960383001044636932034918480\\0629897}     \\
\hline
${\bf A}_9(15,4,5)$ &  \tabincell{l}{147834451659241278745558658029\\146634561} &  \tabincell{l}{147834447219250203363412960695\\716746417 }     \\
\hline
\end{tabular}
\end{table}

\begin{table}[hb]
  \centering
   \caption{Some lower bounds on ${\bf A}_q(4n, 2(n-t), n)$}
  \begin{tabular}{|c|l|l|}\hline
  ${\bf A}_q(n,d,k)$ & Lower Bounds\\
  \hline
  ${\bf A}_2(20,4,5)$ & 1315398998655356311\\
  \hline
  ${\bf A}_3(20,4,5)$ & 43233485281590911580807321041\\
  \hline
  ${\bf A}_4(20,4,5)$ & 1336472440592799231370494712907901631\\
  \hline
  ${\bf A}_5(20,4,5)$ & 869151650599051646738433375279575594407249\\
  \hline
  ${\bf A}_7(20,4,5)$ & 508273020693237561132754855997185401884597574981601 \\
  \hline
  ${\bf A}_8(20,4,5)$ & 1532928970776586688938815376036341347556330253989504511 \\
  \hline
  ${\bf A}_9(20,4,5)$ & 1797321806605534646862867182733878159175088330825288747361 \\
  \hline
  ${\bf A}_2(24,6,6)$ & 4747234173413401936981 \\
  \hline
  ${\bf A}_3(24,6,6)$ & 22530367127371196208130075198509281\\
  \hline
  ${\bf A}_4(24,6,6)$ & 22300867449560834030210344616161360246897891\\
  \hline
  ${\bf A}_5(24,6,6)$ & 211758378832969565256609532806254712000815561347009\\
  \hline
  ${\bf A}_7(24,6,6)$ & 7031676686916460305530685695221278081277908734094305105188801\\
  \hline
  ${\bf A}_8(24,6,6)$ & 105312292581044862467221898491140379101347355113312142905458229671\\
  \hline
  ${\bf A}_9(24,6,6)$ & 507528787550401889216222390017824754036868775577998894410404563393281\\
  \hline
  \end{tabular}
\end{table}

\end{document}